\documentclass[prd,aps,superscriptaddress,nofootinbib,showpacs,showkeys,preprint]{revtex4-1}

\usepackage{amssymb,amsmath,bm}
\usepackage[colorlinks]{hyperref}

\usepackage{graphicx}

\def\PRL#1{{ Phys.\ Rev.\ Lett.} {\bf #1}}
\def\PRD#1{{ Phys.\ Rev.} {\bf D#1}}
\def\NPB#1{{ Nucl.\ Phys.} {\bf B#1}}
\def\PLB#1{{Phys.\ Lett.} {\bf B#1}}

\newcommand{\vev}[1]{\langle #1 \rangle}

\newcommand{\be}{\begin{equation}}
\newcommand{\ee}{\end{equation}}
\newcommand{\bea}{\begin{eqnarray}}
\newcommand{\eea}{\end{eqnarray}}

\begin{document}

\title{Why PeV scale left-right symmetry is a good thing\footnote{\textsl{Presented 
at \textsl{Pheno1} First Workshop on Beyond Standard Model Physics, IISER Mohali April 2016 and at 
the program Exploring the Energy Ladder of the Universe at Mainz Institute for Theoretical Physics 
June 2016}}}

\author{Urjit A. Yajnik}
\affiliation{Physics Department, Indian Institute of Technology Bombay, Mumbai 400076}

\begin{abstract}
Left-right symmetric gauge theory presents a minimal paradigm to accommodate massive 
neutrinos with all known conserved symmetries duly gauged. The work presented here is based on
the argument that the see-saw mechanism does not force the new right handed symmetry scale to be 
very high, and as such some of the species from the spectrum of the new gauge and Higgs bosons can 
have masses within a few orders of magnitude 
of the TeV scale. The scale of the left-right parity breaking in turn can be sequestered from the 
Planck scale by supersymmetry. We have studied several formulations of such Just Beyond Standard 
Model (JBSM) theories for their consistency with cosmology. Specifically the need to eliminate 
phenomenologically undesirable domain walls gives many useful clues. The possibility that the exact 
left-right symmetry breaks in conjunction with supersymmetry has been explored in the context of 
gauge mediation, placing restrictions on the available parameter space. Finally we have also 
studied a left-right symmetric model in the context of metastable supersymmetric vacua and obtained 
constraints on the mass scale of Right handed symmetry. In all the cases studied, The mass scale 
of right handed neutrino $M_R$ remains bounded from above, and in some of the cases the scale 
$10^9$ GeV favourable for supersymmetric thermal leptogenesis is disallowed. On the other hand
PeV scale remains a viable option, and the results warrant a more detailed study of such models for 
their observability in collider and astroparticle experiments. 

\end{abstract}

\keywords      {left-right symmetry, neutrino mass, domain walls, supersymmetry, metastable vacua}

\pacs{12.10.-g,12.60.Jv, 11.27.+d}

\maketitle 

Chirality seems to be an essential feature of fundamental physics,
thereby allowing dynamical generation of fermion masses. However 
this did not require the world to be parity asymmetric, the way it is manifested in
the Standard Model (SM). Indeed, the discovery of neutrino 
masses\cite{Okumura:2016zyh,Fukuda:1998mi}  \cite{Davis:1986ts,solar_data,McDonald:2016xdg} in the 
past two decades 
strongly suggests the existence of right handed neutrino states.
The resulting parity balanced spectrum of fermions begs a parity 
symmetric theory and parity violation could then be explained to
be of dynamical origin. Let us take stock of what principles we could put to use
in a ``down upwards" guesswork in energy scales.
\begin{itemize}
\item Gauge principle provides massless force carriers
\item Chirality provides massless matter
\item Scalars can signal spontaneous symmetry breakdown 
\begin{itemize}
\item provide masses through a universal mechanism
\item  provide naturally substantial amount of $CP$ violation.
\end{itemize}
\item Supersymmetry provides a most comprehensive and elegant way of sequestering mass scales 
signalled by scalars. 
\end{itemize}
Any extension we seek could first be guided by these principles. To be specific we assume 
the restriction that any exact internal symmetry should be gauged, and any scalar and fermionic
fields introduced must be charged under at least one of these gauge symmetries. The left-right
class of models are interesting from this point view and are explored as the first rung of a 
''down upwards'' unification. The two other alternatives not pursued by us are (1) 
$SO(10)$ unification with grand desert, as for instance
proposed in \cite{Poh:2015wta,Aulakh:2002zr,Aulakh:2004hm} or (2) the SM Higgs boson so instrumental
to providing a perturbative description of weak interactions is secretly a member of a strongly 
coupled theory\cite{Lane:2016kvg,Kaul:1981uk}\cite{Csaki:2003si,Kaplan:2003uc,ArkaniHamed:2002qx}.

As baryogenesis began to be ruled out in SM\cite{krs.85,chem_potential_lepto} and leptogenesis 
\cite{fukugita.86} became severely constrained in thermal $SO(10)$ 
scenarios\cite{davidson&ibarra.02} \cite{Boltzmann_equations},
our early observation  \cite{sahu&yaj_prd.04} was that the see-saw mechanism 
\cite{GellMann:1980vs} \cite{Yanagida:1979as} \cite{Mohapatra:1979ia} \cite{Schechter:1980gr}
\cite{Schechter:1981cv} 
generically easily permits an $M_R$ scale as low as $10^6$GeV, considerably smaller than the 
scales of coupling constant unification. Secondly, that low scale non-thermal 
leptogenesis is consistent with the wash out constraints from low scale right-handed neutrinos.  
It is therefore appealing  to look 
for left-right symmetry \cite{leftright_group} as an 
intermediate stage in the sequence of symmetry breaking, and explore the possible range of
masses $M_R$ acceptable for the heavy right handed neutrinos. In the hope that  the new symmetries  
are within the reach of  the LHC and several proposed future colliders, we may call it Just Beyond 
the Standard Model (JBSM).

A number of recent works have focused on this question and studied it in conjunction with 
correlated signatures such as with baryogenesis or leptogenesis, and also additionally Dark Matter 
production 
\cite{Davoudiasl:2015jja,Dev:2015uca,Chala:2016ykx,Kim:2016xyi,Zhuridov:2016xls,Huang:2015izx,
Dhuria:2015xua,Deppisch:2015yqa} and gravitational waves\cite{Huang:2016odd}.
There are also planned or ongoing experiments such as non-accelerator exploration of $N$-$\bar{N}$
oscillations \cite{Frost:2016qzt}, role of right handed currents in the electric dipole moments 
of light nuclei \cite{Dekens:2014jka} and hadronic CP violation \cite{Cirigliano:2016yhc}, 
correlated signatures in 
neutrinoless double 
beta decay \cite{Drewes:2016lqo} \cite{Asaka:2016zib} on the one hand and collider 
signatures\cite{DeSerio:2016ath} \cite{Huang:2015izx} \cite{Djouadi:2007ik} on the other.
Secondly a number of viable scenarios for low scale  leptogenesis exist, such as resonant 
leptogenesis \cite{Pilaftsis:2005rv,Dev:2014tpa} as also the 
generically supersymmetric Affleck-Dine scenario \cite{Affleck:1984fy,Allahverdi:2016yws}. 
We considered specific models \cite{Cline:2002ia} (non-supersymmetric), and \cite{Sarkar:2007er} 
(supersymmetric version) of non-thermal leptogenesis, but it is presumably possible for 
the other scenarios such as referred in this paragraph also 
to be implementable within the parameter space of the left-right models to be considered here. 
Assuming that this is possible within 
the PeV scale, we have studied the consistency of some of the intrinsic  features of the proposed 
left-right symmetric models with cosmology.

\section{Left-right symmetry : a supersymmetric revival}
\label{sec:MSLRM}
Left-right symmetric model\cite{Mohapatra:1980qe, Mohapatra:1979ia} needs a Supersymmetric extension
as an expedient for avoiding the hierarchy problem. 
The minimal set of Higgs superfields required, with
their $SU(3)\otimes$$SU(2)_L\otimes$~$SU(2)_R\otimes$~$U(1)_{\scriptscriptstyle{B-L}}$  is,
\begin{eqnarray}
  {  \Phi_i = ( 1, 2, 2, 0 ),} & \hspace{1em} & {  i  = 1, 2,} \nonumber\\
  {  \Delta = ( 1, 3, 1, 2 ),} & \hspace{1em}  & {   {\Delta}_c = ( 1, 1, 3,  -2 ),}  \nonumber\\
 {  \bar{\Delta} = ( 1, 3, 1, - 2 ),}  &  \hspace{1em}  &   {  \bar{\Delta}_c = ( 1, 1, 3, 2 ),}
 \label{eq:minsulr}
\end{eqnarray}
and further details of the model can be found in the references.

There is an awkward impasse with this model, namely we would like to retain supersymmetry down to
the TeV scale. So the first stage of gauge symmetry breaking has to respect supersymmetry.
If we choose the parameters of the superpotential to ensure spontaneous parity breaking, 
then either the electromagnetic gauge invariance or the $R$ parity have to be 
sacrificed\cite{Kuchimanchi:1993jg,Kuchimanchi:1995vk}. 
The first
of these is unacceptable consequence, and the second entails a requirement of inelegant fixes. 
This problem was elegantly resolved in \cite{Aulakh:1997ba} with further developments in 
\cite{Aulakh:1997fq,Aulakh:1998nn}. 
It contains the two additional triplet Higgs fields 
\begin{equation}
\Omega = ( 1, 3, 1, 0 ), \hspace{2em} \Omega_c = ( 1, 1, 3, 0 )    
\end{equation}
We refer to this briefly as ABMRS model.
Supersymmetric minima breaking $SU(2)_R$ symmetry are signaled by the ansatz
\begin{equation}
  \begin{array}{ccc}
    { \langle \Omega_c \rangle = \left( \begin{array}{cc}
      \omega_c & 0\\
      0 & - \omega_c
    \end{array} \right),} & \hspace{2em} {
    \langle \Delta_c \rangle = \left( \begin{array}{cc}
      0 & 0\\
      d_c & 0
    \end{array} \right),} & 
  \end{array}
\end{equation}
In this model, with an enhanced $R$ symmetry,  we are lead naturally to a see-saw 
relation    $M^2_{B - L} = M_{EW} M_R$.
This means Leptogenesis is postponed to a lower energy scale closer to
$M_{EW}$. Being generically below $10^9$ GeV, this avoids the gravitino mass bound 
but requires non-thermal leptogensis\cite{Sahu:2004sb}.

For comparison we also take an alternative model to this, considered in \cite{Babu:2008ep} 
where a  superfield $S(1,1,1,0)$ also singlet under parity is included
in addition to the minimal set of Higgs of Eq. (\ref{eq:minsulr}). This is referred to here as BM 
model. 

\section{Transitory domain walls}
\label{sec:transitorydomainwalls}
A generic problem long recognised with the model with exact left-right symmetry is that of
cosmological domain walls. The effective potential of the theory is unable to give preference to 
whether $SU(2)_L$ is broken first or $SU(2)_R$. There is therefore a competition to the SM vacuum 
from a 
``parity flipped SM" vacuum, with the low energy effective gauge group 
$SU(3)_c\otimes SU(2)_R\otimes U(1)_{\tilde{Y}}$ where $\tilde{Y}=T_L^3+\frac{1}{2}(B-L)$.
Spontaneous parity breaking 
leads to formation of Domain walls which quickly dominate the energy density of 
the Universe. It is necessary for recovering standard cosmology that these walls
disappear at least before the Big Bang Nucleosynthesis. In an intrinsically 
parity symmetric theory difference in the vacua resulting in destabilisation is 
not permitted. In the following we first obtain the quantitative  requirement on the 
possible asymmetry so that standard cosmology is ensured. In the next two subsections we propose 
mechanisms that could be the source of such asymmetry without causing serious damage to
the basic assumption of exact left-right symmetry.

There are several studies of wall evolution, and an estimate of the temperature at which 
the walls may destabilise,
parametrically expressed in terms of the surface tension of the walls, in turn 
determined by the parity breaking scale $M_R$. By equating
the  terms leading to small symmetry breaking  discussed in the previous para with
this parametric dependence then gives a bound on $M_R$.


The dynamics of the walls in a radiation dominated universe is determined 
by two quantities : \cite{Kibble:1980mv},
\textit{Tension force} $f_T\sim \sigma/R $, where $\sigma$ is energy per 
unit area and $R$ is the average scale of radius of curvature, and
\textit{Friction force} $f_F \sim \beta T^4$ for walls moving with speed
$\beta$ in a medium of temperature $T$. 
The scaling law for 
the growth of the scale $R(t)$ on which the wall  complex is smoothed out,  
is taken to be \( R(t)\approx (G \sigma)^{1/2} t^{3/2} \).
Also, $f_F\sim 1/(Gt^2)$ and $f_T \sim ( \sigma /(G t^3))^{1/2}$. Then 
the pressure difference required to overcome the above forces 
and destabilise the walls is
\begin{equation}
\delta \rho_{\scriptscriptstyle{RD}} \ge G \sigma^2 \approx \frac{M_R^6}{M_{Pl}^2}
\sim M_R^4 \frac{M_R^2}{M_{Pl}^2}
\label{eq:eps-vsix}
\end{equation}

The case of matter dominated evolution is relevant to moduli fields 
copiously produced in generic string inspired models \cite{Kawasaki:2004rx} 
of the Universe.
A wall complex formed at temperature $T_i\sim M_R$ is assumed to have first relaxed 
to being one wall segment per horizon volume. It then becomes comparable in energy 
density  to the ambient matter density, due to the difference in evolution rates,
$1/a(t)$ for walls compared to $1/a^3(t)$ for matter. For simplicity also demand
that the epoch of equality of the two contributions is the epoch also of the onset
of wall destablisation, 
so as to avoid dominance by domain walls. $T_{destab}$ is taken to be the temperature
when the walls begin to be unstable and less dominanat. Thus we can set 
\(
M_{Pl}^{-2} T_{destab}^4 \sim H^2_{eq} \sim \sigma^{\frac{3}{4}} H_i^{\frac{1}{4}} M_{Pl}^{-3}.
\)
The corresponding temperature permits the estimate of the required pressure difference, 
\begin{equation}
 \delta \rho_{\scriptscriptstyle{MD}} > M_R^4 \left(\frac{M_R}{M_{Pl}}\right)^{3/2}
\label{eq:eps-v11half}
\end{equation}
Thus in this case we find $(M_R/M_{Pl})^{3/2}$ \cite{Mishra:2009mk}, a milder suppression
factor than in the radiation dominated case above.

\subsection{Soft terms as a source of lifting parity degeneracy}
\label{sec:csmlgyBrk} 
One source that could provide the required pressure $\delta \rho$ can be sought in a 
generic neutral scalar field $\phi$. It provides the higher 
dimensional operator that may break parity and which has the simple form \cite{Rai:1992xw}
\( 
V_{eff} = \frac{C_5}{M_{Pl}} \phi^5 
\). 
The cosmological requirement then constrains the coefficient $C_5$. In realistic
theories, there are several scalar fields entering such terms, and 
the structure of the latter is conditioned by gauge invariance and supersymmetry.
Here we study a possible source of such asymmetric terms without sacrificing the 
symmetries of the superpotential, viz., in the soft  supersymmetry breaking terms. 
The soft terms which arise in the two models, ABMRS and BM  respectively, 
may be parameterised as follows 
\begin{eqnarray}
\mathcal{L}_{soft}^1 
&=& ~m_1^2 \textrm{Tr} (\Delta \Delta^{\dagger}) +
m_2^2 \textrm{Tr} (\bar{\Delta} \bar{\Delta}^{\dagger})
\nonumber \\
&&  + ~m_3^2 \textrm{Tr} (\Delta_c \Delta^{\dagger}_c) +
m_4^2 \textrm{Tr} (\bar{\Delta}_c \bar{\Delta}^{\dagger}_c)
\label{eq:eqnsoftone}
\\
\mathcal{L}_{soft}^2 
&=&\alpha_1 \textrm{Tr} (\Delta \Omega \Delta^{\dagger}) +
\alpha_2 \textrm{Tr} (\bar{\Delta} \Omega \bar{\Delta}^{\dagger})
\nonumber\\
&&  + ~\alpha_3 \textrm{Tr} (\Delta_c \Omega_c \Delta^{\dagger}_c) +
\alpha_4 \textrm{Tr} (\bar{\Delta}_c \Omega_c \bar{\Delta}^{\dagger}_c)
\label{eq:eqnsofttwo}\\
\mathcal{L}_{soft}^3 
&=& ~\beta_1 \textrm{Tr} (\Omega \Omega^{\dagger}) +
\beta_2 \textrm{Tr} (\Omega_c \Omega^{\dagger}_c)
\label{eq:eqnsoftthree}
\\
\mathcal{L}_{soft}^4 
&=& S[\gamma_1 \textrm{Tr} (\Delta\Delta^{\dagger}) +
\gamma_2 \textrm{Tr} (\bar{\Delta}\bar{\Delta}^{\dagger})]
\nonumber\\
&& + ~ S^*[\gamma_3 \textrm{Tr} (\Delta_c\Delta^{\dagger}_c) +
\gamma_4 \textrm{Tr} (\bar{\Delta}_c\bar{\Delta}^{\dagger}_c)]
\label{eq:eqnsoftfour}
\\
\mathcal{L}_{soft}^5 
&=& ~  {\tilde\sigma}^2 |S|^2
\label{eq:eqnsoftfive} 
\end{eqnarray}
%
For ABMRS model the relevant soft terms are given by,
\begin{equation} 
\mathcal{L}_{soft} = \mathcal{L}_{soft}^1 + \mathcal{L}_{soft}^2
+ \mathcal{L}_{soft}^3
 \end{equation}  
For BM model the soft terms are given by,
\begin{equation} 
\mathcal{L}_{soft} = \mathcal{L}_{soft}^1  + \mathcal{L}_{soft}^4
+ \mathcal{L}_{soft}^5
 \end{equation}  
It can be shown \cite{Preskill:1991kd} that the adequate requirement to evade the domain wall 
problem is to demand $\delta \rho \sim T_{destab}^4$,
where $\delta \rho$ is the difference in the effective potentials across the domain walls and 
$T_{destab}$ is the destablisation temperature, 
at which this difference becomes the dominant force acting on the walls. At a later epoch, the 
walls decay at temperature $T_D$. If the disappearnce of the walls is prompt, the two 
temperatures would be comparable. Even in the case when the epochs are separated in time,
the entropy generated due to final wall disppearance may raise the temprature, bringing it
closer to $T_{destab}$. In the absence of a detailed model we take $T_D \sim T_{destab}$.
In order to obtain the observed Universe we need that 
$T_D$ remains higher than the Big Bang Nucleosynthesis (BBN) scale of a few MeV.
This is the requirement imposed to constrain the 
differences between the soft terms in the Left and Right sectors 
\cite{Sarkar:2007ic, Sarkar:2007er}. In the BM model the
$S$ field does not acquire a vacuum expectation value (vev) in the physically relevant vacua
and hence the terms in eq.s (\ref{eq:eqnsoftfour}) and (\ref{eq:eqnsoftfive}) 
do not contribute to the vacuum energy. The terms in eq. (\ref{eq:eqnsofttwo})
are suppressed in magnitude relative to those in eq. (\ref{eq:eqnsoftthree})
due to having $\Omega$ vev's to one power lower. This argument assumes
that the magnitude of the coefficients $\alpha$ are such as to not mix
up the symmetry breaking scales of the $\Omega$'s and the $\Delta$'s.

To obtain orders of magnitude we have taken the $m_i^2$ parameters 
to be of the form $m_1^2 \sim m_2^2$
$\sim m^2$ and $m_3^2 \sim m_4^2$ $\sim m^{'2}$ \cite{Sarkar:2007er} 
with $T_D$ in the range $ 10 - 10^3$ GeV \cite{Kawasaki:2004rx}. 
For both the models we have taken the value of the $\Delta$ vev's as 
$d \sim 10^4$ GeV. For ABMRS model additionally we take $\omega \sim 10^6 $ GeV.
The resulting differences required for successful removal of domain walls
are shown in Table \ref{tab:DWalls}. 

\begin{table}[bth]
\begin{tabular}{p{.25\textwidth}c|c|c|cp{.3\textwidth}|cp{.3\textwidth}} 
\hline
$T_D/$GeV & $\sim$ & $10$ & $10^2$ & $10^3$ \\ \hline \hline
$(m^2 - m^{\prime 2})/\mbox{GeV}^2$ & $\sim$ &
$10^{-4}  $  & $1$  & $10^{4} $\\ [1mm]
$(\beta_1 - \beta_2)/\mbox{GeV}^2 $ & $\sim$ &
$10^{-8}  $ & $10^{-4} $ & 
$1$ \\ [1mm] \hline \hline
\end{tabular} 
%
\caption{Differences in values of soft supersymmetry breaking parameters 
for a range of domain wall decay temperature values $T_D$. The 
differences signify the extent of parity breaking. 
}
\label{tab:DWalls}
\end{table}
We see from table \ref{tab:DWalls} that assuming both the mass-squared 
differences $m^2-m'^2$ and $\beta_1-\beta_2$ arise from the same dynamics,
$\Omega$ fields are the determinant of the cosmology. This is because
the lower bound on the wall decay temperature $T_D$ required
by $\Omega$ fields is higher and the corresponding $T_D$ is reached 
sooner.  This situation changes if for some reason $\Omega$'s do not
contribute to the pressure difference across the walls. The BM
model does not have $\Omega$'s and falls in this category. 

Between the epoch of destabilisation of the DW and their
decay, leptogenesis occurs due to preferential motion of the DW to 
select SM as the low energy theory, as discussed in
\cite{Cline:2002ia, Sarkar:2007er}. After the disappearance, i.e., complete decay 
of the walls at the scale $T_D$, electroweak symmetry breaks at a scale $M_{EW} \sim 10^2$ 
GeV and standard cosmology takes over.

Since the soft terms are associated with supersymmetry breaking, they may be assumed
to arise from the same mechanism that breaks supersymmetry.  In Sec. 
\ref{sec:customgmsb} we discuss the implementation of gauge mediated supersymmetry 
breaking (GMSB) scenario for these models and treat the soft terms to have  arisen 
from the hidden sector and communicated
along with the messenger fields \cite{Mishra:2008be}. Constraints on the hidden 
sector model and the communication mechanism can be obtained in this way.

\subsection{Parity breaking from Planck suppressed effects}
The soft terms studied above are tied to the scale of supersymmetry breaking.
Another mechanism to look into, without incurring violence to the symmetries of the 
superpotential, is to assume that the parity breaking operators arise at the Planck scale
\cite{Mishra:2009mk}. Writing the lowest order terms in the superpotential suppressed 
by the Planck scale, and using the K\"{a}hler potential 
formalism, we obtain the expectation value of the effective potential of the scalar fields 
after substituting the vacuum expectation values of the relevant fields as follows,  
whose details can be found in \cite{Mishra:2009mk}.
\begin{equation}
  V^R_{eff} \sim \frac{a(c_R + d_R)}{M_{Pl}}M_R^4M_W + \frac{a(a_R + d_R)}{M_{Pl}}
M_R^3M_W^2
\end{equation}
and likewise $R\leftrightarrow L$. Here the constants $a$, $c_R$ etc. are dimensionless.
Hence, with regrouping the above coefficients into coefficients $\kappa$ etc.,
which for naturalness should remain order unity,
\begin{equation}
  \delta\rho \sim \kappa^A \frac{M_R^4M_W}{M_{Pl}}+{\kappa'}^A\frac{M_R^3M_W^2}{M_{Pl}}
\end{equation}
It is interesting to study this relation with $\kappa$ coefficients $O(1)$ and $\delta\rho \sim 
(10\mathrm{MeV})^4$, compatible with BBN. It then leads without any further assumptions
to a lower bound on $M_R$ to be same as SM scale, leaving all higher scales open to being
physically viable.

Now inserting these answers in the expectations derived from cosmological dynamics of DW, 
viz., $\delta \rho_{\scriptscriptstyle{RD}}$, $ \delta \rho_{\scriptscriptstyle{MD}}$
derived in  Eq.s (\ref{eq:eps-vsix}) and (\ref{eq:eps-v11half}),
we obtain
\begin{equation}
 \kappa^A_{RD} > 10^{-10} \left( \frac{M_R}{10^6 {\rm GeV}}\right)^2
\end{equation}
Thus only minuscule values of $\kappa$ coefficients suffice to allow PeV scale $M_R$.
At the higher end, $M_R$ $\lesssim 10^9$GeV  needed to avoid gravitino problem after 
reheating at the end of inflation remains viable with $\kappa_{RD}\sim 10^{-4}$,
and also $M_R$ of the order of the intermediate scale $10^{11}$GeV remains marginally viable 
with $\kappa^A_{RD}$ $\sim O(1)$. But scales higher than that are not tolerated by naturalness.

Next, if we consider wall disappearance during moduli dominated regime, we find
\begin{equation}
 \kappa^A_{MD} > 10^{-2} \left( \frac{M_R}{10^6 {\rm GeV}}\right)^{3/2},
\end{equation}
which can be seen to be a rather strict requirement barely allowing $M_R$ higher than
our preferred PeV scale. In particular, taking $M_R\sim10^9$GeV 
required to have thermal leptogenesis without the undesirable gravitino  production, leads to 
unnatural values of $\kappa_{MD}>10^{5/2}$.

\section{Customised GMSB for Left-Right symmetric models} 
\label{sec:customgmsb}
The differences required between the soft terms of the Left and the
Right sector for the DW to disappear at a temperature $T_D$ 
are not unnaturally large. However the reasons for appearance of 
even a small asymmetry between the Left and the Right fields is hard to
explain since the theory as adopted in Sec. \ref{sec:MSLRM} is parity symmetric. 
We now try to explain the origin of this small difference by
focusing on the hidden sector, and relating it to supersymmetry (SUSY) breaking.

For this purpose we assume that the strong dynamics responsible for SUSY
breaking also breaks parity, which is then transmitted to the visible sector
via the messenger sector and is encoded in the soft supersymmetry breaking terms.
We implement this idea by introducing two singlet fields $X$ and $X'$, 
respectively even and odd under parity.
\begin{equation}
X \leftrightarrow X, \qquad X' \leftrightarrow -X'.
\end{equation} 
The messenger sector superpotential then contains terms
\begin{eqnarray}
W &=& \sum_n \left[ \lambda_n X \left( \Phi_{nL} \bar{\Phi}_{nL} 
+ \Phi_{nR} \bar{\Phi}_{nR}\right) \right.
\nonumber \\ &&
+ \left. ~\lambda'_n X' \left(\Phi_{nL} \bar{\Phi}_{nL} 
- \Phi_{nR} \bar{\Phi}_{nR} \right) \right]
\end{eqnarray} 
For simplicity, we consider $n=1$. The fields $\Phi_{L}$, $\bar\Phi_{L}$
and $\Phi_{R}$, $\bar\Phi_{R}$ are complete representations of
a simple gauge group embedding the L-R symmetry group. Further we
require that the fields labelled $L$ get exchanged with fields
labelled $R$ under an inner automorphism which exchanges
$SU(2)_L$ and $SU(2)_R$ charges, e.g. the charge conjugation operation 
in $SO(10)$. As a simple possibility we consider the case when
$\Phi_{L}$, $\bar\Phi_{L}$ (respectively, $\Phi_{R}$, $\bar\Phi_{R}$) 
are neutral  under $SU(2)_R$ ($SU(2)_L$). Generalisation to other 
representations is straightforward.

As a result of the dynamical SUSY breaking we expect the fields
$X$ and $X'$ to develop nontrivial vev's and $F$ terms and hence
give rise to mass scales
\begin{equation}
\Lambda_X = \frac{\vev{F_X}}{\vev{X}},
\qquad
\Lambda_{X'} = \frac{\vev{F_{X'}}}{\vev{X'}}.
\end{equation} 
Both of these are related to the dynamical SUSY breaking scale $M_S$,
however their values are different unless additional reasons of symmetry
would force them to be identical. Assuming that they are different
but comparable in magnitude we can show that Left-Right
breaking can be achieved simultaneously with SUSY breaking being
communicated. 

In the proposed model, the messenger fermions receive respective mass 
contributions
\begin{eqnarray} 
m_{f_L} &=& |\lambda\vev{X} + \lambda^{\prime}\vev{X^{\prime}}|\\ \nonumber
m_{f_R} &=& |\lambda\vev{X} - \lambda^{\prime}\vev{X^{\prime}}|
 \end{eqnarray} 
while the messenger scalars develop the masses
\begin{eqnarray}
m_{\phi_L}^2 &=&  |\lambda\vev{X} + \lambda^{\prime}\vev{X^{\prime}}|^2
\pm  |\lambda\vev{F_X} + \lambda^{\prime}\vev{F_{X^{\prime}}}| \\ \nonumber
m_{\phi_R}^2 &=&  |\lambda\vev{X} - \lambda^{\prime}\vev{X^{\prime}}|^2
\pm  |\lambda\vev{F_X} - \lambda^{\prime}\vev{F_{X^{\prime}}}|
\end{eqnarray}
We thus have both SUSY and parity breaking communicated through these
particles.

As a result the mass contributions to the gauginos 
of $SU(2)_L$ and $SU(2)_R$ from  both the $X$  and $X'$ fields with 
their corresponding auxiliary parts take the simple form,
\begin{equation}
M_{a_{L}} = \frac{\alpha_a}{4 \pi} 
\frac{\lambda \langle F_X \rangle + \lambda^{\prime}
\langle F_{X^{\prime}} \rangle}{\lambda \langle X \rangle
+ \lambda^{\prime} \langle X^{\prime} \rangle}
\end{equation}
and
\begin{equation}
M_{a_{R}} = \frac{\alpha_a}{4 \pi}
\frac{\lambda \langle F_X \rangle - \lambda^{\prime}
\langle F_{X^{\prime}} \rangle}{\lambda \langle X \rangle
- \lambda^{\prime} \langle X^{\prime} \rangle} 
\end{equation}
upto terms suppressed by $\sim F/X^2$.
Here $a = 1, 2, 3$. 
In turn there is a modification to  
scalar masses, through two-loop corrections, expressed to leading
orders in the $x_L$ or $x_R$ respectively, by  the generic formulae
\begin{eqnarray}
m^2_{\phi_{L}} &=&
2 \left( \frac{\lambda \langle F_X \rangle + \lambda^{\prime}
\langle F_{X^{\prime}} \rangle}{\lambda \langle X \rangle
+ \lambda^{\prime} \langle X^{\prime} \rangle}\right)^2\\
&\times& \left [ \left (\frac{\alpha_3}{4\pi}\right )^2 C_3^\phi +
\left (\frac{\alpha_2}{4 \pi}\right )^2
(C_{2L}^\phi)
 + \left (\frac{\alpha_1}{4 \pi}\right )^2 C_1^\phi \right ]
\label{eq:ModSclrMs1} 
\end{eqnarray}

\begin{eqnarray}
m^2_{\phi_{R}} &=&
2 \left( \frac{\lambda \langle F_X \rangle - \lambda^{\prime}
\langle F_{X^{\prime}} \rangle}{\lambda \langle X \rangle
- \lambda^{\prime} \langle X^{\prime} \rangle}\right)^2\\
&\times&\left [ \left (\frac{\alpha_3}{4\pi}\right )^2 C_3^\phi +
\left (\frac{\alpha_2}{4 \pi}\right )^2
(C_{2R}^\phi)
 + \left (\frac{\alpha_1}{4 \pi}\right )^2 C_1^\phi \right ]
\label{eq:ModSclrMs2} 
\end{eqnarray}

The resulting difference 
between the mass squared of the left and right 
sectors are obtained as
\begin{equation}
\delta m_\Delta^2 
=
2 (\Lambda_X)^2 f(\gamma, \sigma)
\left\{  \left(\frac{\alpha_2}{4\pi}\right)^2
+ \frac{6}{5} \left( \frac{\alpha_1}{4 \pi} \right)^2 \right\}
\label{eq:Dm2Delta}
\end{equation}
where,
\begin{equation}
 f(\gamma, \sigma) = 
\left( \frac{ 1 + \textrm{tan}\gamma}{1+
\textrm{tan} \sigma}\right)^2
- \left( \frac{ 1 - \textrm{tan} \gamma}{ 1 -
 \textrm{tan} \sigma}\right)^2 
\label{eq:tanfnc}
\end{equation}
We have brought $\Lambda_X$ out as the representative mass scale
and parameterised the ratio of mass scales by introducing 
\begin{equation}
\textrm{tan}\gamma = \frac{\lambda^{\prime} 
\langle F_{X^{\prime}} \rangle}{\lambda \langle F_X \rangle},
\quad \textrm{tan} \sigma = \frac{\lambda^{\prime} 
\langle X^{\prime}\rangle}{\lambda \langle X \rangle}
\end{equation}
Similarly, 
\begin{equation}
\delta m_\Omega^2 = 2 (\Lambda_X)^2 
f(\gamma, \sigma)
\left( \frac{\alpha_2}{4\pi} \right)^2
\label{eq:Dm2Omega}
\end{equation}
In the models studied here, the ABMRS model will have contribution
from both the above kind of terms. The BM model will have contribution
only from the $\Delta$ fields.

The contribution to slepton masses is also obtained from eq.s (\ref{eq:ModSclrMs1})
and (\ref{eq:ModSclrMs2}).
This can be used to estimate the magnitude of the overall scale $\Lambda_X$
to be $\geq 30$ TeV based on \cite{Dubovsky:1999xc} from LEP limits, which may not
have changed significantly even in the light of LHC data 
\cite{Knapen:2016exe,Allanach:2016pam}. 
Substituting this in the above formulae (\ref{eq:Dm2Delta}) and (\ref{eq:Dm2Omega})
we obtain the magnitude of the factor $f(\gamma, \sigma)$ required for
cosmology as estimated in table \ref{tab:DWalls}. The resulting values of 
$f(\gamma, \sigma)$ are tabulated in table \ref{tab:lambdasp}. We see that the natural 
range of temperature for the disappearance 
of domain walls therefore remains TeV or higher, i.e., upto a few order of magnitudes 
lower  than the  scale at which they form. 

\begin{table}
\begin{tabular}{cc|c|c|c} 
\hline
$T_D/$GeV & $\sim$ & $10$ & $10^2$ & $10^3$ \\ \hline \hline
Adequate $(m^2-m'^2)$ & $\phantom{\sim}$
& $ 10^{-7} $  & $ 10^{-3} $  & $10$\\ 
Adequate $(\beta_1-\beta_2)$  & $\phantom{\sim}$
& $10^{-11}$ & $ 10^{-7}$ & $10^{-3}$ \\ [1mm] \hline \hline
\end{tabular}
\caption{Entries in this table are the values of the parameter $f(\gamma,\sigma)$, 
required to ensure  wall decay at temperature $T_D$
displayed in the header row. The table
should be read in conjunction with table \ref{tab:DWalls},
with the rows corresponding to each other.}
\label{tab:lambdasp}
\end{table}

\section{Supersymmetry breaking in metastable vacua}
The dilemma of phenomenology with broken supersymmetry can be captured in the  
fate of $R$ symmetry generic to superpotentials \cite{Nelson:1993nf}. 
An unbroken $R$ symmetry in the theory is required for SUSY breaking. 
$R$ symmetry when spontaneously broken leads  to $R$-axions which are 
unacceptable. If we give up $R$ symmetry, the ground state remains supersymmetric.
The solution proposed in \cite{Nelson:1993nf}, is to 
break $R$ symmetry  mildly, governed by a small parameter $\epsilon$.
Supersymmetric vacuum persists, but this can be pushed far away in
field space.  SUSY breaking local minimum is ensured near the origin, 
since it persists in the limit $\epsilon\rightarrow 0$.
A specific example of this scenario \cite{Seiberg:1994pq,Intriligator:2006dd,Intriligator:2007py}  
referred to as ISS, 
envisages $SU(N_c)$ SQCD with $N_f>N_c$ flavours of quarks $q, \tilde{q}$ which is UV free, such 
that 
it is dual to a $SU(N_f-N_c)$ gauge theory which is IR free, the so called
magnetic phase, with $N^2_f$ singlet mesons $M$.

Thus we consider a Left-Right symmetric model with ISS mechanism
as proposed in \cite{Haba:2011pr}. They proposed the electric gauge theory to be based on the gauge 
group $SU(3)_L \times SU(3)_R \times SU(2)_L \times SU(2)_R \times U(1)_{B-L}$ 
(in short $G_{33221}$) where $SU(2)_L \times SU(2)_R \times U(1)_{B-L}$ is the gauge group 
of usual Left-Right models and $SU(3)_{L,R}$ is the new strongly coupled gauge sector introduced. 
The dual description similar to the original ISS model gives rise to $SU(2)_R$ broken meta-stable 
vacua inducing spontaneous SUSY breaking simultaneously. 

The particle content of the electric theory is
$$ Q_L^a \sim (3,1,2, 1, 1), \quad \tilde{Q}^a_L \sim (3^*, 1, 2, 1, -1) $$
$$ Q_R^a \sim (1,3,1, 2, -1), \quad \tilde{Q}^a_R \sim (1,3^*,1, 2, 1) $$
where $a = 1, N_f$ and the numbers in brackets correspond to the transformations of the fields 
under 
the gauge group $G_{33221}$. This model has $N_c = 3$ and hence to have a Seiberg dual 
\cite{Seiberg:1994pq} magnetic theory, number of flavours should be $N_f \geq 4$. For $N_f = 4$ the 
dual magnetic theory will have the gauge symmetry of the usual Left Right Models $SU(2)_L \times 
SU(2)_R \times U(1)_{B-L}$ and the following particle content
$$\phi^a_L (2,1,-1), \quad \tilde{\phi}^a_L (2,1,1) $$
 $$\phi^a_R (1,2,1), \quad \tilde{\phi}^a_R (1,2,-1) $$
\begin{equation}
\Phi_L \equiv \bf{1}+\text{Adj}_L =
\left(\begin{array}{cc}
\ \frac{1}{\surd 2}(S_L+\delta^0_L) & \delta^+_L \\
\ \delta^-_L & \frac{1}{\surd 2}(S_L-\delta^0_L )
\end{array}\right) \nonumber
\end{equation}
\begin{equation}
\Phi_R \equiv \bf{1}+\text{Adj}_R =
\left(\begin{array}{cc}
\ \frac{1}{\surd 2}(S_R+\delta^0_R) & \delta^+_R \\
\ \delta^-_R & \frac{1}{\surd 2}(S_R-\delta^0_R )
\end{array}\right)
\label{eq:deltaCcomponents}
\end{equation}
The Left-Right symmetric renormalisable superpotential of this
magnetic theory is 
\begin{equation}
W^0_{LR} = h \mathrm{Tr} \phi_L \Phi_L \tilde{\phi}_L -h \mu^2 \mathrm{Tr}
\Phi_L+h \mathrm{Tr} \phi_R \Phi_R \tilde{\phi}_R -h \mu^2 \mathrm{Tr}
\Phi_R
\label{WLR1}
\end{equation}
After integrating out the right handed chiral fields, the
superpotential becomes
\begin{equation}
W^0_{L} = h \mathrm{Tr} \phi_L \Phi_L \tilde{\phi}_L -h \mu^2 \mathrm{Tr}
\Phi_L+ h^4 \Lambda^{-1} \mathrm{det}\Phi_R -h \mu^2 \mathrm{Tr} \Phi_R
\label{WL1}
\end{equation}
which gives rise to SUSY preserving vacua at 
\begin{equation}
\langle h \Phi_R \rangle = \Lambda_m \epsilon^{2/3} = \mu
\frac{1}{\epsilon^{1/3}}
\label{rightvev}
\end{equation}
where $\epsilon = \frac{\mu}{\Lambda_m}$.
Thus the right handed sector exists in a metastable SUSY breaking
vacuum whereas the left handed sector is in a SUSY preserving vacuum
breaking D-parity spontaneously.

We next consider \cite{Borah:2011aw} Planck scale suppressed terms that may signal parity breaking
\begin{eqnarray} 
W^1_{LR}& =& f_L \frac{\mathrm{Tr}(\phi_L \Phi_L \tilde{\phi}_L) \mathrm{Tr}
\Phi_L}{\Lambda_m} +f_R \frac{\mathrm{Tr}(\phi_R \Phi_R \tilde{\phi}_R)
\mathrm{Tr} \Phi_R}{\Lambda_m}\nonumber \\ 
&&+ f'_L \frac{(\mathrm{Tr} \Phi_L)^4}{\Lambda_m}+f'_R \frac{(\mathrm{Tr} \Phi_R )^4}{\Lambda_m}
\end{eqnarray}
 The terms of order $\frac{1}{\Lambda_m}$ are given by
\begin{eqnarray}
V^1_R &=&  \frac{h}{\Lambda_m} S_R [f_R(\phi^0_R
\tilde{\phi}^0_R)^2+f'_R\phi^0_R \tilde{\phi}^0_R S^2_R \nonumber \\
&+&(\delta^0_R
-S_R)^2((\phi^0_R)^2+(\tilde{\phi}^0_R)^2)]
\end{eqnarray}

The minimisation conditions give $\phi \tilde{\phi} = \mu^2 $ and $S^0
= -\delta^0$. Denoting $\langle \phi^0_R \rangle = \langle
\tilde{\phi}^0_R \rangle = \mu $ and $\langle \delta^0_R \rangle
=-\langle S^0_R \rangle = M_R$, we have 
\begin{equation}
V^1_R = \frac{hf_R}{\Lambda_m} (\lvert \mu \rvert^4 M_R +\lvert \mu
\rvert^2 M^3_R ) 
\end{equation}
where we have also assumed $f'_R \approx f_R$. For $ \lvert \mu \rvert
< M_R$ 
Thus the effective energy density difference between the two types of vacua is
\begin{equation}
\delta \rho \sim h(f_R-f_L) \frac{\lvert \mu \rvert^2
M^3_R}{\Lambda_m}
\end{equation}
Now we set $\mu\sim$TeV, the scale of supersymmetry breaking and since QCD cannot 
be expected to break parity, assume that the parity breaking terms are essentially 
of Planck scale origin. Using this self consistent requirement,
for walls disappearing in matter dominated era, we get
\begin{equation}
M_R < \lvert \mu \rvert^{5/9} M^{4/9}_{Pl} \sim 1.3 \times 10^{10}\; \mathrm{GeV}
\end{equation}
with $\mu\sim$TeV.
Similarly for the walls disappearing in radiation dominated era,
\begin{equation}
M_R < \lvert \mu \rvert^{10/21} M^{11/21}_{Pl} \sim 10^{11}\; \mathrm{GeV}
\end{equation}
Thus in both cases we get an upper bound on the $M_R$ which can at best be an intermediate scale,
while the PeV scale remains eminently viable.

\section{Conclusions}
We have pursued the possibility of left-right symmetric models as Just Beyond Standard Models 
(JBSM), not possessing a large hierarchy.  We also adopt the natural points of view that right 
handed neutrinos must be included in the JBSM with local gauge symmetric interactions and that 
the required 
parity breaking to match low energy physics arises from spontaneous breakdown.
The latter scenario is often eschewed due to the domain walls it entails in the early 
Universe. We turn the question around to ask given that the domain walls occur,
what physics could be responsible for their successful removal without jeopardising
naturalness. 

Seeking origins of parity breaking that are consistent with the principles outlined in the 
Introduction, we proceed to correlate this breaking to supersymmetry breaking. 
We have considered
three models along these lines. One involves gauge mediated supersymmetry breaking, another 
assumes Planck scale breaking and the third relies on metastable vacua for supersymmetry breaking.
 The operators permissible in these scenarios are then
constrained by the cosmological requirements on the dynamics of domain wall disappearance.
In all the cases studied, The mass scale 
of right handed neutrino $M_R$ remains bounded from above, and in some of the cases the scale 
$10^9$ GeV favourable for supersymmetric thermal leptogenesis is disallowed. On the other hand
PeV scale remains a viable option.

The possibility that such a low energy model maybe embedded successfully in the semisimple 
$SO(10)$ at a high scale has been explored separately \cite{Kopp:2009xt}\cite{Borah:2010kk}.
The general message seems to be that the parity breaking scale is not
warranted to be as high as the grand unification scale and further,
several scenarios suggest that left-right symmetry as the Just Beyond Standard Model 
package incorporating the SM may be within the reach of future colliders.

\section{Acknowledgment}
This is a comprehensive resume of the work done jointly with Narendra Sahu, Anjishnu Sarkar, 
Sasmita Mishra and Debasish Borah. It is a happy duty to thank the collaborators. The coverage of 
other works along similar lines suffers from the 
limitation of space permitted here but also from limitation of my awareness of them. Most of the  
work was supported by grants from Department of Science and Technology. I thank the organisers of 
the 
Pheno1 conference IISER Mohali and the organisers and participants of the program Exploring the 
Energy Ladder of the Universe for enabling presentation and discussion of this work. The author 
would like to express a special thanks to the Mainz Institute for Theoretical Physics (MITP) for 
its hospitality and support. 

\bibliographystyle{pramana}

\end{document}